\documentclass[11pt,a4paper]{article}

\usepackage[utf8]{inputenc}
\usepackage[T1]{fontenc}
\usepackage[left=2.5cm, right=2.5cm, top=2.5cm, bottom=2.5cm]{geometry}
\usepackage{amsmath, amssymb, amsthm}
\usepackage{graphicx}
\usepackage{booktabs}
\usepackage{tabularx}
\usepackage[justification=raggedright, singlelinecheck=false, skip=0pt]{caption}
\usepackage{titlesec}
\usepackage{enumitem}
\usepackage{xcolor}
\usepackage{authblk}
\usepackage{lineno}
\usepackage{float}
\usepackage{fancyhdr}
\usepackage[hyphens]{url}
\usepackage{makecell}

\renewenvironment{abstract}
  {\noindent\small\textbf{Abstract:}\enspace}
  {\par} 

\newcommand{\keywords}[1]{
  \noindent\begin{flushleft}\small\textbf{Keywords:}\enspace #1\end{flushleft}
  \nointerlineskip
  \noindent\rule{\textwidth}{0.8pt}
  \vskip 2em
}

\makeatletter
\renewcommand{\maketitle}{
  \begin{flushleft}
    \nointerlineskip
    \noindent\rule{\textwidth}{0.8pt}
    {\huge \bfseries \@title \par}
    \vskip 2em
    {\@author \par}
    \vskip 1.5em
  \end{flushleft}
}

\makeatother

\renewenvironment{abstract}
  {\noindent\small\textbf{Abstract:}\enspace}
  {\par\vskip 1em}

\theoremstyle{plain}

\theoremstyle{definition}

\titleformat{\section}{\normalfont\large\bfseries}{\thesection.}{0.5em}{}
\titlespacing{\section}{0pt}{3.5ex plus 1ex minus .2ex}{0.5ex}

\titleformat{\subsection}{\normalfont\normalsize\itshape}{\thesubsection.}{0.5em}{}
\titlespacing{\subsection}{0pt}{2ex plus 0.5ex minus .2ex}{0.1ex}

\titleformat{\subsubsection}{\normalfont\normalsize}{\thesubsubsection.}{0.5em}{}
\titlespacing{\subsubsection}{0pt}{0.5ex plus 0.2ex minus .1ex}{0.1ex}

\title{GPCR Ligand Bioactivity Prediction with Physics-Informed Dual-State Query Learning}
\author{\textbf{Shuo Zhang}}
\author{\textbf{Huifeng Zhang}}
\author{\textbf{Rongqi Hong}}
\author[1,*]{\textbf{Jian K. Liu}}
\affil{University of Birmingham}
\affil[*]{Correspondence: j.liu.22@bham.ac.uk}
\date{}

\begin{document}

\maketitle
\thispagestyle{fancy}

\begin{abstract}
Predicting the bioactivity profiles of small molecules against G protein-coupled receptors (GPCRs) is a challenge in drug discovery. Although deep learning has accelerated the prediction of binding affinities, existing approaches often struggle to distinguish between functional efficacies because they neglect dynamic conformational equilibria. Furthermore, structure-based methods are frequently limited by the scarcity of high-resolution active-state crystal structures and the indistinguishability of conformational states in static representations. To bridge the gap between black-box prediction and biophysical reality, we propose Dual-State Query (DSQ), a physics-informed multimodal architecture that explicitly embeds the Monod-Wyman-Changeux (MWC) model of allostery within a neural network. Unlike conventional models that rely on explicit 3D structures, DSQ utilizes learnable orthogonal queries to extract disentangled representations of active and inactive receptor states. These latent representations are governed by a novel neural MWC gating module, which mathematically derives the probability of receptor activation from thermodynamic competition between ligand-state affinities and the receptor's intrinsic conformational energy barrier. A contrastive ranking objective is also introduced to enforce differential affinity constraints, ensuring physical consistency. Extensive experiments demonstrate that DSQ outperforms other baselines, particularly for the agonist subset. Additional homology-stratified, temperature-sensitivity, perturbation, clustering, and efficiency analyses show that DSQ provides useful thermodynamic inductive bias, while also exposing a clear limitation on low-homology receptors. The code is available at https://github.com/jiankliu/DSQ.
\end{abstract}

\keywords{binding affinity prediction; learnable query; GPCR}

\section{Introduction}
G protein-coupled receptors (GPCRs) represent the largest superfamily of membrane proteins in the human genome and serve as primary targets for a substantial portion of FDA-approved drugs~\cite{Zhang2024gprotein,Conflitti2025functional}. The therapeutic potential of a GPCR ligand is determined not only by its binding affinity but fundamentally by its functional effect: whether it activates (agonist) or inhibits (antagonist) the receptor. This pharmacological property is governed by the receptor's ability to transition between different active and inactive conformational states~\cite{Kobilka2007conformational}. Consequently, accurate GPCR profiling requires a dual prediction capability: quantifying binding strength while simultaneously characterizing the functional state induced by the ligand.

Despite its importance, integrating structural mechanisms into data-driven prediction remains a significant challenge. Traditional structure-based drug design (SBDD) relies on high-resolution crystal structures to model interactions~\cite{Congreve2020impacr}. However, active-state GPCR structures are scarce due to their inherent instability~\cite{Carpenter2016engineering}. Even when structures are available, the high geometric similarity between active and inactive states, particularly in the transmembrane region, often confounds standard geometric deep learning models. Structure-based prediction is also limited when reliable binding pockets or active-state conformations are unavailable. While recent hierarchical architectures have attempted to capture distinct energetic signals at flexible structural boundaries or domain interfaces~\cite{zhang2026dagml}, they still heavily rely on explicit 3D conformations. Consequently, these models remain constrained when applied to GPCRs, where active-state crystals are scarce and dynamic conformational equilibria cannot be fully resolved by static geometric graphs. In contrast, ligand-based quantitative structure-activity relationship (QSAR) models typically neglect the receptor's structural context~\cite{Thomas2021comparison}. Generalization to receptor-disjoint or orphan GPCRs therefore remains a distinct challenge that requires dedicated evaluation.

In this work, we propose Dual-State Query (DSQ), a physics-informed multimodal architecture that explicitly embeds the Monod-Wyman-Changeux (MWC) model of allostery~\cite{Canals2012monod} into a deep learning framework. Unlike conventional models that represent proteins as static embeddings, DSQ introduces a latent state probing mechanism. We utilize learnable orthogonal state queries to extract disentangled structural representations of active and inactive states from a pre-trained protein language model (ESM-2~\cite{Lin2023esm2}), circumventing the need for explicit 3D structures. These representations are governed by a novel neural MWC gating module, which mathematically derives the probability of receptor activation from the thermodynamic competition between ligand-state affinities and the receptor's intrinsic energy barrier. By enforcing differential affinity constraints through a contrastive ranking objective, DSQ ensures that predictions are thermodynamically consistent.

Our contributions are summarized as follows:
\begin{itemize}[topsep=0pt, partopsep=0pt, parsep=0pt]
    \item Physics-Informed Architecture: We propose DSQ, a novel framework that embeds the MWC allosteric model into a neural network. By simulating the receptor's conformational ensemble via learnable state queries, we address the limitations of static structural representations.
    \item Joint Affinity-Efficacy Prediction: DSQ uses efficacy labels during training but, at inference, predicts activation probability and bioactivity jointly without requiring an externally supplied agonist/antagonist class token.
    \item Comprehensive Evaluation: We validate DSQ on a large-scale human GPCR benchmark curated from GLASS and GPCRdb. Beyond standard test-set performance, we add homology-stratified evaluation, temperature sensitivity, perturbation-based motif validation, quantitative clustering metrics, and computational profiling. These analyses support the benefit of the MWC inductive bias while identifying low-homology generalization as an unresolved limitation.
\end{itemize}

\section{Problem Definition}
Formally, let $\mathcal{P}$ denote the space of protein amino acid sequences and $\mathcal{L}$ denote the space of ligand molecular graphs. Given a pair $(p,l)\in \mathcal{P}\times \mathcal{L}$, the task of GPCR bioactivity profiling is to simultaneously predict the binding affinity $\hat{y}\in \mathbb{R}$ and the functional efficacy class $\hat{c}\in{0,1}$ (where 1 denotes agonist and 0 denotes antagonist). We denote the DSQ model by $\Phi_{\theta}:\mathcal{P}\times\mathcal{L}\rightarrow \mathbb{R}\times[0,1]$, where $\theta$ denotes all trainable model parameters.

Unlike standard affinity prediction tasks, which target $\hat{y}$ in isolation, we formulate this as a physics-constrained learning problem governed by the MWC model. We assume the receptor exists in a thermodynamic equilibrium between an active state ($R^*$) and an inactive state ($R$). The observed efficacy $\hat{c}$ is modeled as the outcome of the activation probability $P(R^*|p,l)$, a latent variable derived from the ligand's differential affinity toward these two states. Thus, we optimize $\theta$ so that $\hat{y}$ minimizes the regression objective while the latent $P(R^*|p,l)$ is consistent with the functional label.

\begin{figure*}[h]
  \centering
  \includegraphics[width=\linewidth]{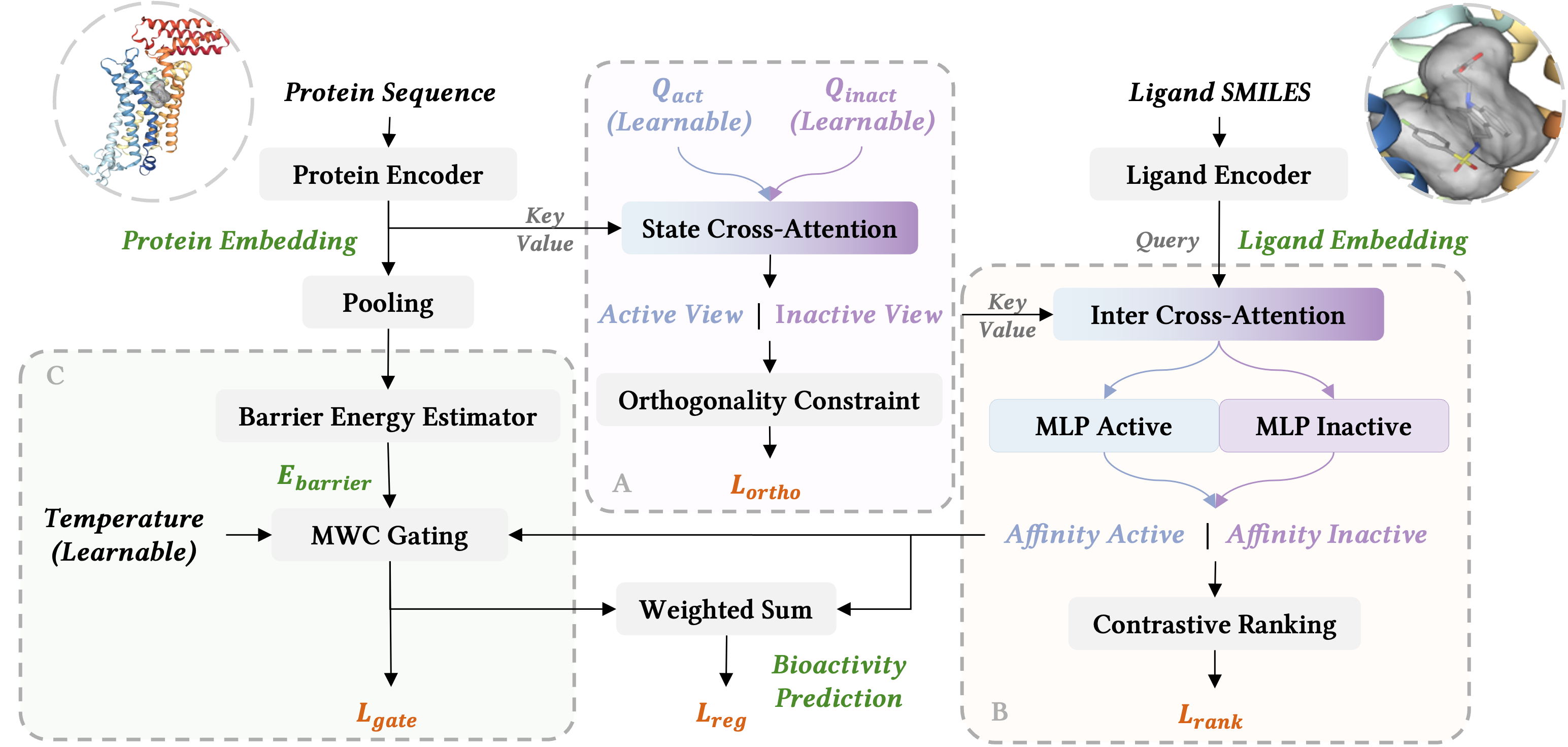}
  \caption{Model Architecture: A. Latent state probing; B. Interaction learning; C. Neural MWC module. The parameters of State and Inter Cross-Attention layers are shared between two state-specific query branches.}
  \label{fig:model_arch}
\end{figure*}

\section{Materials and Methods}
\label{sec:method}

\subsection{Overview}
Dual-State Query (DSQ) simulates the conformational selection of GPCRs in a latent space. As illustrated in Figure~\ref{fig:model_arch}, the architecture comprises three main components: A. Latent state probing, which uses learnable queries to extract disentangled active and inactive state representations; B. Interaction learning, which models ligand binding to each state; and C. a Neural MWC module, which enforces thermodynamic consistency between affinities and activation probability.

\subsection{Input Representation}

We use general-purpose pretrained encoders rather than GPCR-specific foundation models. For protein $p$, ESM-2 (esm2\_t33\_650M\_UR50D) provides residue-level embeddings $H_P^{(0)}\in\mathbb{R}^{N\times1280}$. For ligand $l$, MoLFormer-c3-1.1B~\cite{Ross2022molformer} provides token-level embeddings $H_L^{(0)}\in\mathbb{R}^{M\times d_l}$. ESM-2 provides general protein sequence representations, whereas MolFormer provides general chemical representations. ESM-2 features are precomputed and MolFormer parameters are frozen during DSQ training. Learned projections map both modalities to the common hidden dimension $d$:
\begin{equation}
\tilde{H}_P=H_P^{(0)}W_P,\qquad
\tilde{H}_L=H_L^{(0)}W_L.
\end{equation}

\subsection{Latent State Probing via Learnable Queries}
We introduce a latent probe mechanism to approximate receptor conformational views in the embedding space. The operation is implemented as cross-attention pooling; the contribution is not a new attention primitive, but the use of two state-specific probe sets whose outputs are coupled to an MWC-inspired thermodynamic gate. We initialize two sets of learnable query vectors: $Q_{act}\in \mathbb{R}^{K\times d}$ for the active state and $Q_{inact}\in \mathbb{R}^{K\times d}$ for the inactive state. Here, $K$ represents the number of latent structural probes and $d$ denotes the hidden dimension. These queries interact with the protein embeddings through multi-head cross-attention:

\begin{equation}
    Z_s=\operatorname{MHA}(Q_s,\tilde{H}_P,\tilde{H}_P),
    \qquad s\in\{act,inact\},
\end{equation}
where $Z_s$ represents the latent structural view of state $s$.

\subsection{Interaction Learning and Intrinsic Affinity}
Ligand features $H_L$ are fused with the state-specific protein representations ($Z_{act}$, $Z_{inact}$) via a second cross-attention layer, where $H_L$ is query and $Z_s$ is key/value. The resulting interaction embeddings are pooled and projected by state-specific MLPs to predict the intrinsic affinities $\hat{y}_{act}$ and $\hat{y}_{inact}$:
\begin{align}
    \hat{y}_{act}&=\text{MLP}_{act}(\text{Interaction}(H_L,Z_{act})) \\
    \hat{y}_{inact}&=\text{MLP}_{inact}(\text{Interaction}(H_L,Z_{inact}))
\end{align}
Here, $\hat{y}_{act}$ and $\hat{y}_{inact}$ are latent state-specific affinity estimates constrained by the physics-informed gate described below. The two prediction heads have identical architectures but independent parameters. Each head uses linear layers with dimensions $d\rightarrow d\rightarrow d/2\rightarrow1$, with batch normalization, GELU activation, and dropout of 0.1 after the first hidden layer.

\subsection{Neural MWC Gating Mechanism}
The DSQ architecture integrates the MWC biophysical model into a differentiable neural layer. The MWC model posits that receptor activation is driven by the thermodynamic stability of the ligand-receptor complex in the active state relative to the inactive state.

\subsubsection{Theoretical Derivation}
Let $G_{act}$ and $G_{inact}$ be the Gibbs free energies of the ligand-bound receptor in the active and inactive states, respectively. Following the Boltzmann distribution, the probability of the receptor being in the active state $R^*$ is given by:
\begin{equation}
\begin{split}
    P(R^*)=\frac{e^{-\beta G_{act}}}{e^{-\beta G_{act}}+e^{-\beta G_{inact}}}=\sigma(\beta(G_{inact}-G_{act}))
\end{split} 
\end{equation}
where $\beta=1/k_BT$, $k_B$ is the Boltzmann constant, $T$ is the absolute temperature, and $\sigma(\cdot)$ is the sigmoid function. 

The binding affinity $y$ (e.g., $pK_i$) is linearly proportional to the negative binding free energy:
\begin{equation}
    y\propto-\Delta G_{bind}=k_BT\text{ln}K_a
\end{equation}
where $K_a$ represents the association constant. 

According to the MWC model, the total free energy difference can be decomposed into the intrinsic conformational energy barrier of the receptor ($E_{barrier}$) and the change in free energy due to ligand interaction ($\Delta G_{bind}$). We can thus rewrite the energy difference as:
\begin{equation}
    G_{inact}-G_{act}\propto(y_{act}-y_{inact})-E_{barrier}
\end{equation}
where $y_{act}-y_{inact}$ represents the ligand's preferential stabilization of the active state, and $E_{barrier}$ represents the basal energetic cost for the receptor to transition to the active state in the absence of a ligand~\cite{Deupi2010energy}.

\subsubsection{Neural Implementation}
We model $E_{barrier}(p)$ using a projection head on the pooled protein embedding. The activation probability is then computed as follows:
\begin{equation}
    P(R^*|p,l)=\sigma\left(\frac{(\hat{y}_{act}-\hat{y}_{inact})-E_{barrier}(p)}{\tau}\right)
\end{equation}
where $\tau$ is a learnable temperature parameter initialized to 1.0. It calibrates the scale of latent affinity differences before the sigmoid gate. This mechanism enforces physical logic: to predict agonism ($P(R^*)\rightarrow1$), the model must learn that the ligand preferentially stabilizes the active state relative to the inactive state ($\hat{y}_{act}\gg\hat{y}_{inact}$) and overcomes the learned conformational offset $E_{barrier}(p)$. In contrast, antagonists are modeled as ligands that either preferentially stabilize the inactive state or fail to stabilize the active state sufficiently.

The final predicted bioactivity is the weighted expectation of the state affinities:
\begin{align}
    \hat{y}_{final}=P(R^*|p,l)\cdot\hat{y}_{act}+(1-P(R^*|p,l))\cdot\hat{y}_{inact}
\end{align}

\subsection{Physics-Informed Optimization}
To ensure that the learned representation adheres to the thermodynamic constraints posited by the MWC model, we optimize a composite objective function consisting of four terms:
\begin{enumerate}
    \item Regression Loss $\mathcal{L}_{reg}$: Standard mean squared error between $\hat{y}_{final}$ and experimental ground truth.
    \item MWC Gate Supervision $\mathcal{L}_{gate}$: Supervises the latent activation probability $P(R^*|p,l)$ using binary cross-entropy against the functional labels (1 for agonists, 0 for antagonists), forcing the latent state distribution to align with pharmacological efficacy.
    \item Orthogonality Constraint $\mathcal{L}_{ortho}$: Enforces distinct structural views for active and inactive states by minimizing their cosine similarity:
    \begin{equation}
        \mathcal{L}_{ortho}=\mathbb{E}\left[\frac{|Z_{act}\cdot Z_{inact}|}{\left\| Z_{act}\right\|_2 \cdot \left\| Z_{inact}\right\|_2}\right]
    \end{equation}
    \item Contrastive Ranking Loss $\mathcal{L}_{rank}$: Enforces differential affinity consistent with efficacy labels using a margin ranking loss ($\hat{y}_{act}>\hat{y}_{inact}$ for agonists ($y_{label}=1$); $\hat{y}_{inact}>\hat{y}_{act}$ for antagonists):
    \begin{equation}
        \mathcal{L}_{rank}=\text{max}(0, -r\cdot(\hat{y}_{act}-\hat{y}_{inact})+m)
    \end{equation}
    where $r=2\cdot y_{label}-1$ converts the binary label to {-1,1}, and $m$ is a margin hyperparameter.
\end{enumerate}

The total loss is a weighted sum: 
\begin{equation}
    \mathcal{L}_{total}=\lambda_1\mathcal{L}_{reg}+\lambda_2\mathcal{L}_{gate}+\lambda_3\mathcal{L}_{ortho}+\lambda_4\mathcal{L}_{rank},
\end{equation}
where $\lambda_1=1.0$ and $\lambda_2=\lambda_3=\lambda_4=0.5$. The ranking margin is $m=1.0$.

This optimization scheme guides the model to learn a physics-based manifold where bioactivity predictions emerge from the interplay of distinct conformational states.

\subsection{Dataset Construction and Setup}
Following the dataset curation protocol of AiGPro~\cite{Brahma2025aigpro}, we constructed the benchmark from GLASS~\cite{xu2025glass2,Chan2015glass} and GPCRdb~\cite{Herrera2024gpcrdb}. We retained $IC_{50}$ and $K_i$ records as antagonist bioactivities and $EC_{50}$ records as agonist bioactivities. All bioactivity values ($IC_{50}, EC_{50}, K_i$) were standardized to the negative logarithmic scale with nM unit ($pX=9-\text{log}_{10}X$) to facilitate regression stability.

We adopted the independent test set released by AiGPro, containing 11,464 interaction pairs across 203 receptors. Test interaction pairs were removed before model training, and the remaining data were randomly divided into training (80\%) and validation (20\%) sets. Table~\ref{tab:dataset} summarizes the statistics of the final data. The resulting training set comprises 260,816 samples covering 222 unique GPCRs, with a notable class imbalance reflecting real-world drug discovery scenarios (approximately 3.3:1 antagonist-to-agonist ratio).

\begin{table}
\centering
\caption{Dataset Statistics}
\label{tab:dataset}
\begin{tabular}{l|cccc}
\toprule
Split       & Antagonist    & Agonist   & Total Samples     & Unique Proteins   \\
\midrule
Train       & 200,258       & 60,558 	& 260,816 	       & 222   \\
Val         & 49,674        & 15,531  	& 65,205 	       & 218   \\
Test        & 5,095         & 6,369 	& 11,464 	       & 203  \\
\bottomrule
\end{tabular}
\end{table}

\textbf{Implementation Details:} The model was implemented in PyTorch and optimized with Adam using a learning rate of $5\times10^{-4}$. ESM-2 and MolFormer were frozen, leaving 3.96M trainable parameters in DSQ. Training ran for up to 100 epochs on a single 40GB NVIDIA A100 GPU with early stopping patience of 5. The observed training cost was approximately 474 seconds per epoch on the A100. Inference throughput was approximately 470 samples/s on the NVIDIA RTX 4090.

\section{Results}
We evaluated DSQ against DeepDTAGen~\cite{Shah2025deepdtagen}, DrugBAN~\cite{Bai2023drugban}, and GraphDTA~\cite{Nguyen2020graphdta}. All reported baselines were retrained on the same train/validation/test interaction split and evaluated using the same bioactivity targets and metrics. Each baseline retained its native architecture and input representation.

As shown in Table~\ref{tab:main_res}, DSQ achieves the best overall result for every reported metric among the evaluated models. Relative to the strongest baseline for each metric, DSQ reduces RMSE from 1.071 to 0.999 (6.7\%) and MAE from 0.816 to 0.746 (8.6\%), while improving $R^2$ from 0.446 to 0.517 (15.9\%), PCC from 0.700 to 0.728 (4.0\%), SCC from 0.671 to 0.718 (7.0\%), and C-Index from 0.783 to 0.820 (4.7\%).

The improvement is concentrated in the overall and agonist results. DSQ does not uniformly outperform the baselines on the antagonist subset; DeepDTAGen and GraphDTA achieve better results on several antagonist-specific metrics. We therefore interpret the principal benefit of DSQ as improved modeling of the more difficult agonist subset rather than a universal improvement in antagonist affinity prediction.

One challenge in GPCR profiling is the data imbalance between agonists (active) and antagonists (inactive), with agonists typically being the minority class and exhibiting more restricted conformational compatibility. This is evidenced by the baseline performance on the agonist subset: DeepDTAGen and GraphDTA yield negative $R^2$ values (-0.084 and -0.282, respectively), indicating that their predictions for agonists are worse than a simple mean baseline. DrugBAN shows marginal correlation ($R^2$=0.099). In contrast, DSQ achieves an $R^2$ of 0.269 on the agonist subset. This result suggests that explicitly modeling active and inactive latent states helps DSQ recover the agonist-specific signal that is weakened when the task is treated as a single scalar regression problem.

\begin{table}
\centering
\caption{Performance comparison on GPCR bioactivity profiling dataset. Metrics are reported for the overall test set and stratified by pharmacological classification (Agonist vs. Antagonist).}
\label{tab:main_res}
\begin{tabular}{l|ccccccc}
\toprule
Model       & Type      & RMSE$\downarrow$      & MAE$\downarrow$       & $R^2$$\uparrow$    & PCC$\uparrow$       & SCC$\uparrow$       & C-Index$\uparrow$    \\
\midrule
DeepDTAGen  & Overall   & 1.078 	& 0.860 	& 0.438 	& 0.700 	& 0.671 	& 0.783 \\
            & Agonist   & 1.078 	& 0.894 	& -0.084 	& 0.589 	& 0.431 	& 0.656 \\
            & Antagonist & 1.078 	& 0.817 	& 0.417 	& 0.650 	& 0.655 	& 0.737 \\ \hline
DrugBAN     & Overall   & 1.071 	& 0.816 	& 0.446 	& 0.670 	& 0.609 	& 0.748 \\
            & Agonist   & 0.983 	& 0.782 	& 0.099 	& 0.552 	& 0.341 	& 0.621 \\
            & Antagonist & 1.171 	& 0.858 	& 0.312 	& 0.641 	& 0.650 	& 0.741 \\ \hline
GraphDTA    & Overall   & 1.150 	& 0.909 	& 0.361 	& 0.628 	& 0.594 	& 0.726 \\
            & Agonist   & 1.173 	& 0.975 	& -0.282 	& 0.511 	& 0.445 	& 0.659 \\
            & Antagonist & 1.120 	& 0.827 	& 0.370 	& 0.649 	& 0.662 	& 0.741 \\
\midrule
DSQ         & Overall   & \textbf{0.999}     & \textbf{0.746} 	& \textbf{0.517} 	& \textbf{0.728} 	& \textbf{0.718} 	& \textbf{0.820} \\
            & Agonist   & 0.886 	& 0.668 	& 0.269 	& 0.643 	& 0.477 	& 0.671 	\\
            & Antagonist & 1.125 	& 0.845 	& 0.364 	& 0.620 	& 0.624 	& 0.723 	\\
\bottomrule
\end{tabular}
\end{table}

\subsection{Homology-Stratified Evaluation}
To examine whether performance is driven by receptor homology, we stratified the independent test set by the maximum sequence identity between each test receptor and the training receptors. Table~\ref{tab:homology_res} reports high-homology ($\geq$80\%) and low-homology ($<$60\%) subsets.

\begin{table*}[htbp]
\centering
\caption{Homology-stratified performance. High: test receptors with maximum sequence identity $\geq$80\% to training receptors. Low: test receptors with maximum sequence identity $<$60\%.}
\label{tab:homology_res}
\begin{tabular}{ll|ccc}
\toprule
Model & Subset & RMSE$\downarrow$ & $R^2$$\uparrow$ & C-Index$\uparrow$ \\
\midrule
DSQ & High Overall & 0.910 & 0.584 & 0.777 \\
    & High Agonist & 0.838 & 0.263 & 0.671 \\
    & High Antagonist & 1.010 & 0.492 & 0.768 \\
    & Low Overall & 1.472 & -0.081 & 0.578 \\
    & Low Agonist & 1.573 & -0.095 & 0.555 \\
    & Low Antagonist & 1.442 & -0.093 & 0.578 \\
\midrule
DeepDTAGen & High Overall & 1.023 & 0.474 & 0.754 \\
    & High Agonist & 1.062 & -0.184 & 0.651 \\
    & High Antagonist & 0.962 & 0.539 & 0.781 \\
    & Low Overall & 1.388 & 0.039 & 0.604 \\
    & Low Agonist & 1.376 & 0.162 & 0.637 \\
    & Low Antagonist & 1.392 & -0.018 & 0.588 \\
\midrule
GraphDTA & High Overall & 1.079 & 0.415 & 0.742 \\
    & High Agonist & 1.153 & -0.395 & 0.664 \\
    & High Antagonist & 0.955 & 0.546 & 0.799 \\
    & Low Overall & 1.553 & -0.203 & 0.535 \\
    & Low Agonist & 1.536 & -0.044 & 0.561 \\
    & Low Antagonist & 1.558 & -0.276 & 0.527 \\
\bottomrule
\end{tabular}
\end{table*}

DSQ performs best on the high-homology overall and agonist subsets, improving high-homology overall $R^2$ over DeepDTAGen and GraphDTA. However, the low-homology results show a clear generalization boundary: all models degrade to near-zero or negative $R^2$. These results indicate that the current sequence-based formulation remains limited when test receptors cross large evolutionary distances from the training receptors.

\subsection{Ablation Study}
To validate the necessity of the physics-informed constraints proposed in Section~\ref{sec:method}, we conducted an ablation study (Table~\ref{tab:component_res}).

\textbf{Effect of the MWC Gating Mechanism}: Replacing the MWC gating mechanism with a standard MLP ("simple gate") results in a performance drop across all metrics (RMSE increases from 0.999 to 1.051) and a notable decrease in gate accuracy (0.840 to 0.747). Because this variant preserves a comparable gating role but removes the MWC parameterization, the drop suggests that the gain is not explained solely by adding model capacity. Instead, the thermodynamic relationship between differential affinity and activation probability contributes to the improvement.

\textbf{Effect of Learnable Temperature}: In the full DSQ model, $\tau$ decreased from 1.0 to 0.2627, indicating that the model learned to sharpen the activation gate during training. Freezing $\tau$ at 1.0 reduces overall $R^2$ from 0.517 to 0.427 and gate accuracy from 0.840 to 0.812. The effect is larger on agonists, where $R^2$ drops from 0.269 to 0.142 and SCC drops from 0.477 to 0.386. These results indicate that learned temperature calibration improves activation-state discrimination, especially for the minority agonist subset.

\begin{table}[!htbp]
\centering
\small
\setlength{\tabcolsep}{5pt}
\caption{Ablation study of DSQ components. simple gate: replaces the MWC gate net with MLP module; $\tau$=1.0: freezes the temperature parameter; w/o gate: treats the task as pure regression without state disentanglement; w/o rank: removes contrastive ranking loss; w/o ortho: removes orthogonality constraint. Gate Acc: accuracy of classifying agonist vs. antagonist.} 
\label{tab:component_res}
\begin{tabular}{l|cccccccc}
\toprule
Model       & Type      & RMSE      & MAE       & $R^2$     & PCC       & SCC       & C-Index   & \makecell{Gate \\ Acc}  \\
\midrule
DSQ         & Overall   & 0.999     & 0.746 	& 0.517 	& 0.728 	& 0.718 	& 0.820     & 0.840     \\
            & Agonist   & 0.886 	& 0.668 	& 0.269 	& 0.643 	& 0.477 	& 0.671 	&          \\
            & Antagonist & 1.125 	& 0.845 	& 0.364 	& 0.620 	& 0.624 	& 0.723 	&          \\\midrule
$\tau$=1.0  & Overall   & 1.089     & 0.810 	& 0.427 	& 0.674 	& 0.676 	& 0.805     & 0.812     \\
            & Agonist   & 0.960 	& 0.728 	& 0.142 	& 0.584 	& 0.386 	& 0.640 	&          \\
            & Antagonist & 1.231 	& 0.912 	& 0.239 	& 0.539 	& 0.571 	& 0.703 	&          \\\midrule
simple gate & Overall   & 1.051 	& 0.780 	& 0.466 	& 0.698 	& 0.690 	& 0.806 	& 0.747  \\
            & Agonist   & 0.928 	& 0.694 	& 0.198 	& 0.620 	& 0.446 	& 0.664     & \\
            & Antagonist & 1.187 	& 0.888 	& 0.292 	& 0.582 	& 0.593 	& 0.713     &  \\ \hline
w/o rank    & Overall   & 1.114 	& 0.857 	& 0.400 	& 0.659 	& 0.637 	& 0.782 	& 0.745 \\
            & Agonist   & 1.028 	& 0.811 	& 0.015 	& 0.569 	& 0.315 	& 0.613     & \\
            & Antagonist & 1.213 	& 0.914 	& 0.261 	& 0.544 	& 0.561 	& 0.700     & \\ \hline
w/o ortho   & Overall   & 1.141 	& 0.895 	& 0.371 	& 0.627 	& 0.605 	& 0.780 	& 0.672  \\
            & Agonist   & 1.019 	& 0.804 	& 0.031 	& 0.525 	& 0.243 	& 0.588 	& \\
            & Antagonist & 1.277 	& 1.008 	& 0.182 	& 0.461 	& 0.441 	& 0.654 	& \\ \hline
w/o gate    & Overall   & 1.159 	& 0.867 	& 0.351 	& 0.622 	& 0.601 	& 0.750 	& 0.532 \\
            & Agonist   & 1.079 	& 0.821 	& -0.086 	& 0.541 	& 0.394 	& 0.645     & \\
            & Antagonist & 1.251 	& 0.925 	& 0.215 	& 0.558 	& 0.568 	& 0.706     & \\
\bottomrule
\end{tabular}
\end{table}

\textbf{Effect of Contrastive Ranking}: Removing the ranking loss ("w/o rank") impacts the model's ability to distinguish preferential binding to the active state, with the active $R^2$ dropping to near zero (0.015). Although the overall regression performance remains moderate, the gate accuracy drops to 0.745. This confirms that explicit enforcement of $\hat{y}_{act}>\hat{y}_{inact}$ of agonists is required to align latent affinity estimates with thermodynamic constraints.

\textbf{Effect of Orthogonality Constraint}: Removing the orthogonality constraint ("w/o ortho") leads to a degradation in RMSE to 1.141 and gate accuracy to 0.672. This suggests that the query vectors likely suffer from mode collapse, learning identical features rather than disentangled conformational states.

\textbf{Necessity of the Gating Mechanism}: The "w/o gate" variant represents a standard regression model using the same backbone. It shows the lowest performance and near-random classification accuracy (0.532). Under this benchmark, the explicit gating formulation provides a stronger inductive bias than the corresponding scalar-regression ablation.

\subsection{Verification of the Neural MWC Mechanism}
We investigated whether DSQ adhered to the physical principles of the MWC model, specifically regarding differential affinity and conformational energy barriers.

\begin{figure}[h]
  \centering
  \includegraphics[width=0.7\linewidth]{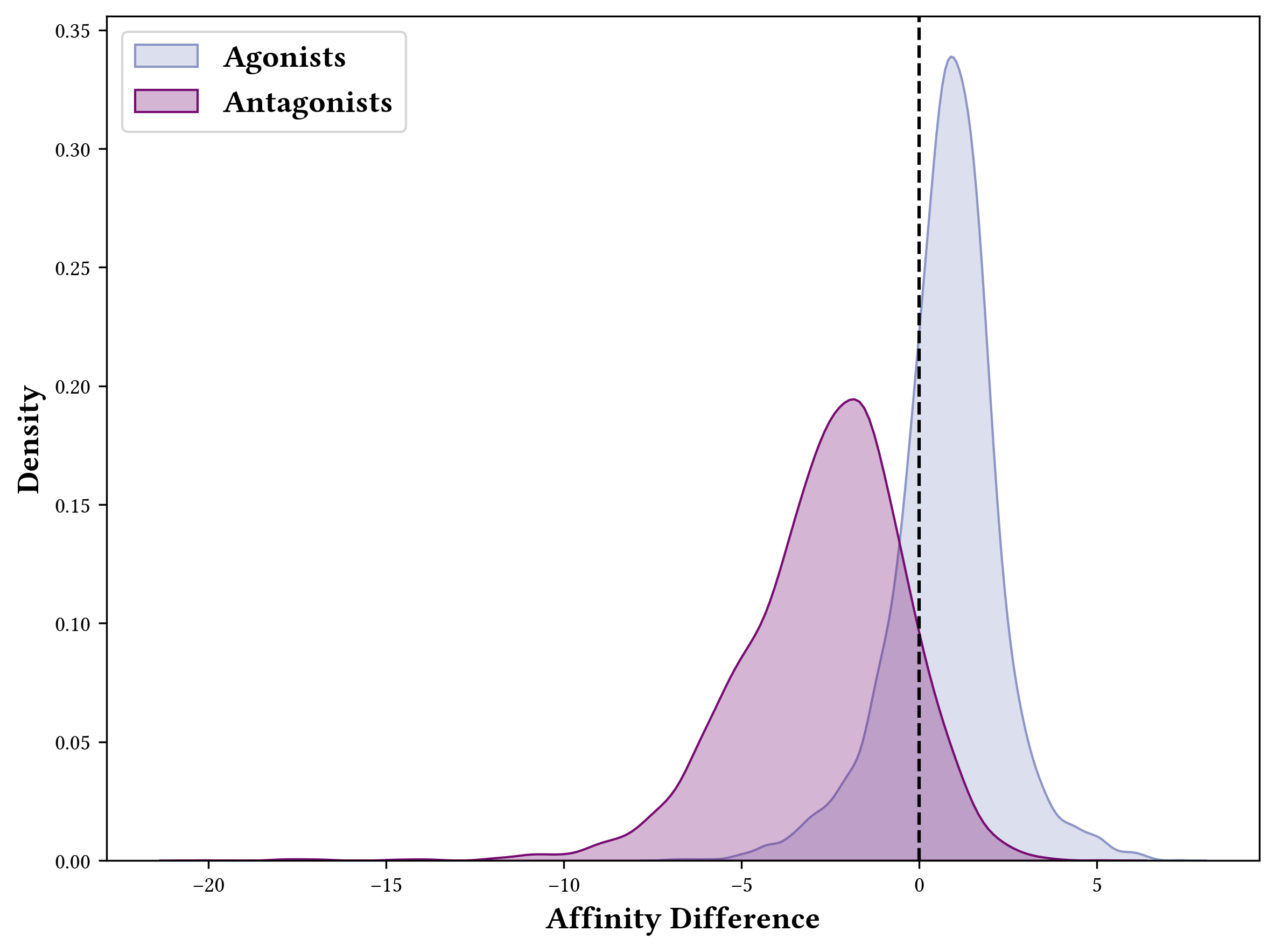}
  \caption{Distribution of predicted differential affinity.}
  \label{fig:diff_affinity}
\end{figure}

\textbf{Differential Affinity Analysis}. We computed the differential affinity $\Delta\hat{y}=\hat{y}_{act}-\hat{y}_{inact}$ for all ligand-receptor pairs in the test set. As shown in Figure~\ref{fig:diff_affinity}, the distribution of $\Delta\hat{y}$ for agonists is shifted to the right compared to antagonists. Quantitatively, the model exhibits a high degree of thermodynamic consistency: 74.91\% of the agonists were predicted to have a higher affinity for the active state, and 90.17\% of the antagonists were predicted to have a higher affinity for the inactive state.

\textbf{Conformational Energy Analysis}. We further analysed the learned conformational energy barrier $E_{barrier}$, which quantifies the intrinsic energetic cost for a receptor to transition to the active state in the absence of a ligand. Predicted $E_{barrier}$ values (mean -0.0481, std 0.1881) could be interpreted as latent offsets. The nonzero variance indicates that DSQ assigns receptor-dependent activation offsets instead of using a single global threshold. Thus, the model calibrates activation probability according to protein sequence context.

\subsection{Interpretability of Learned Queries: Rediscovering Biological Switches}
We use the latent probe attention maps as hypothesis-generating signals and further test selected motifs through perturbation. We computed differential attention ($\Delta A=|A_{active}-A_{inactive}|$) to identify residues where the model attends differently for active versus inactive state predictions.

\begin{figure}[t]
  \centering
  \includegraphics[width=\linewidth]{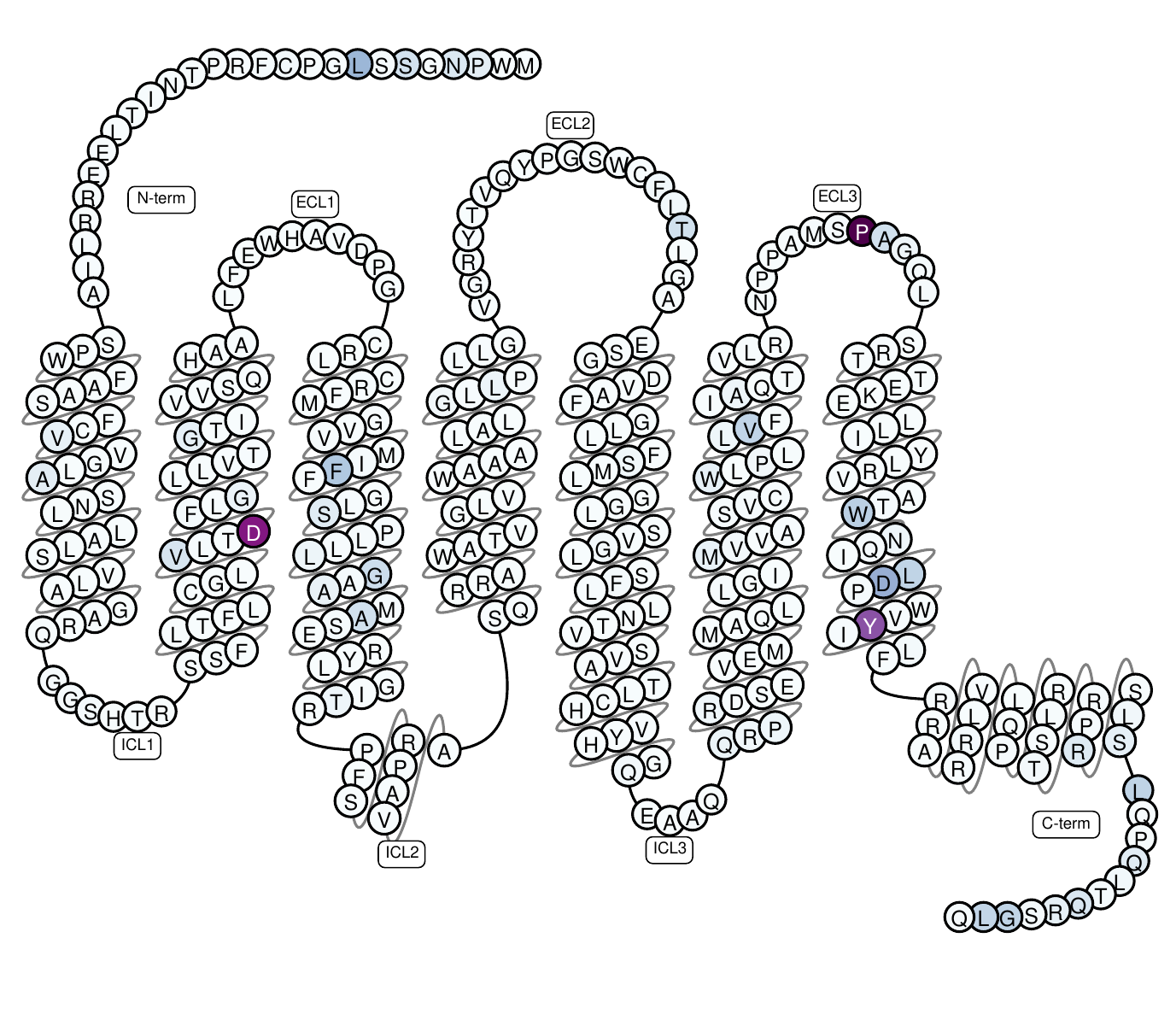}
  \caption{Weighted snake-plot of human thromboxane $A_2$ receptor}
  \label{fig:ta2r_snakeplot}
\end{figure}

\begin{figure}
  \centering
  \includegraphics[width=0.7\linewidth]{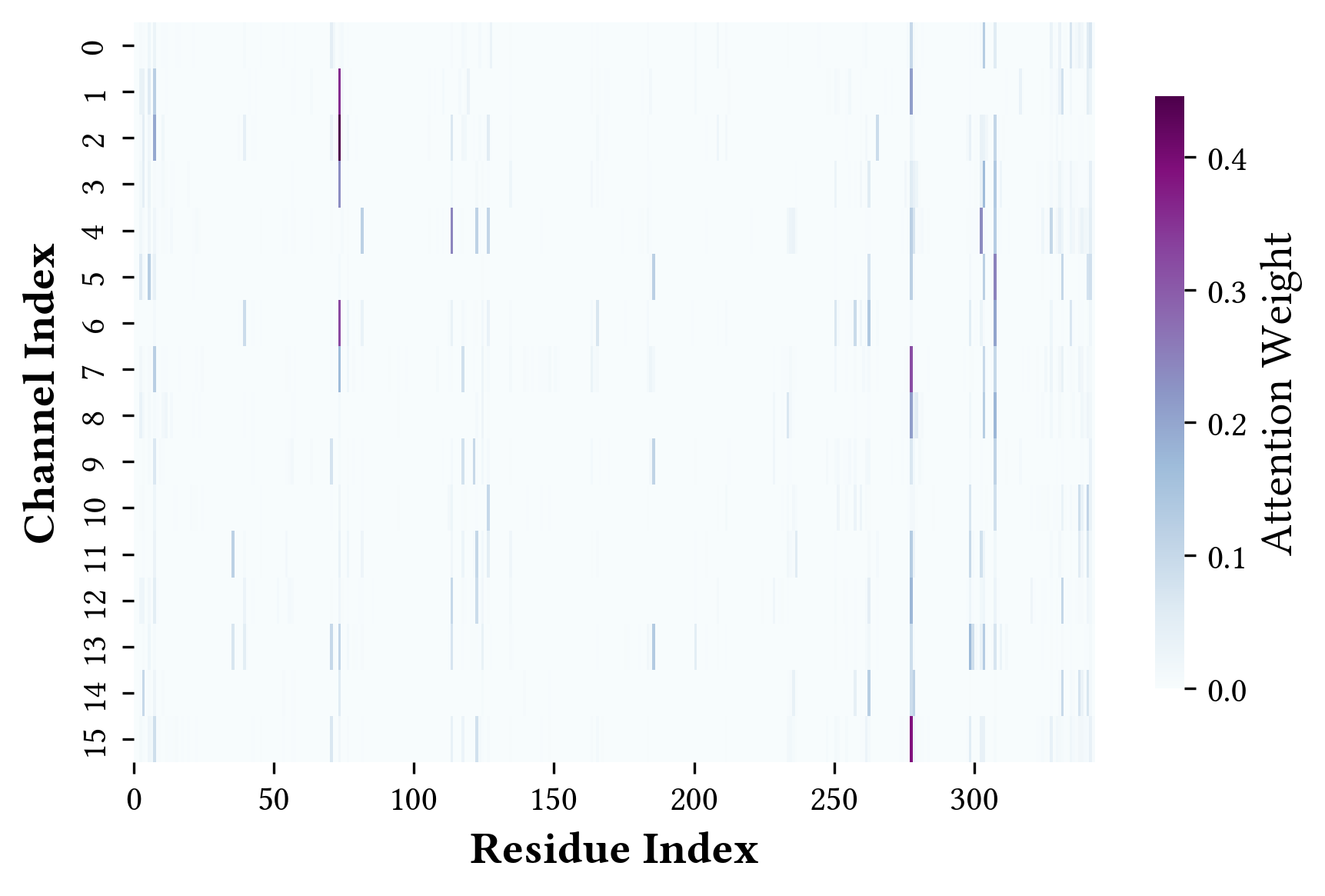}
  \caption{Per-channel attention weights of learnable queries on human thromboxane $A_2$ receptor}
  \label{fig:ta2r_perchannel}
\end{figure}

\textbf{Class A Activation Determinants}. We first examined the human thromboxane $A_2$ receptor (UniProt: P21731) as a representative Class A GPCR. The mean differential attention map reveals statistically significant peaks ($z\text{-score} > 4$) at several key regulatory sites (Figure~\ref{fig:ta2r_snakeplot}). High $\Delta A$ is observed at position 2x50, a highly conserved residue known to form the sodium ion binding pocket, an allosteric site for stabilizing the inactive state in Class A GPCRs~\cite{Hauser2021gpcractivation,Madsen2022mechanistic}. Furthermore, significant attention is directed towards 7x53 (NPxxY motif), a microswitch essential for G-protein coupling and receptor activation~\cite{Murali2024divergent,Zhou2019common}. We also observed high attention in ECL3, a region involved in ligand entry and selectivity~\cite{Zhang2024gprotein}. Analysing each query channel reveals that different heads specialize in different structural motifs (Figure~\ref{fig:ta2r_perchannel}), suggesting that the model learns a distributed representation of the activation landscape.

Expanding this analysis to the entire Class A in test set via multiple sequence alignment confirms the generality of these findings (Figure~\ref{fig:attn_weight_classA_heatmap}). Significant differential attention patterns are consistently observed at 2x50 and 7x53. The model also attends to the N-terminus and ECL2, regions important for ligand recognition and initial binding events~\cite{Wheatley2012lifting}.

\textbf{Class B1 Activation Determinants}. For Class B1 GPCRs (Figure ~\ref{fig:attn_weight_classB1_heatmap}), significant differential attention is observed at 2x50, part of the conserved HETx motif, which stabilizes the inactive state and is disrupted upon activation~\cite{Cary2022newinsights}. We also find high attentions at 7x53 and 7x54 in the NPxxY motif. Furthermore, the model identifies 8x51 in Helix 8, a region known to interact with G-proteins and regulate surface expression~\cite{Huynh2009role}.

\textbf{Perturbation Validation}. To test whether these highlighted motifs affect model output, we performed in silico feature knockout on the human thromboxane $A_2$ receptor by zeroing the embeddings of the sodium pocket residue 2x50 and the NPxxY motif positions 7x49-7x53. For agonists ($N=46$), the mean predicted activation probability $P(R^*)$ decreased from 0.5147 to 0.4294. For antagonists ($N=1204$), $P(R^*)$ remained low, changing from 0.0496 to 0.0393. This perturbation result supports the relevance of these motifs for agonist activation predictions.

\subsection{Interpretability of Latent Space}
\begin{figure*}[h]
  \centering
  \includegraphics[width=\linewidth]{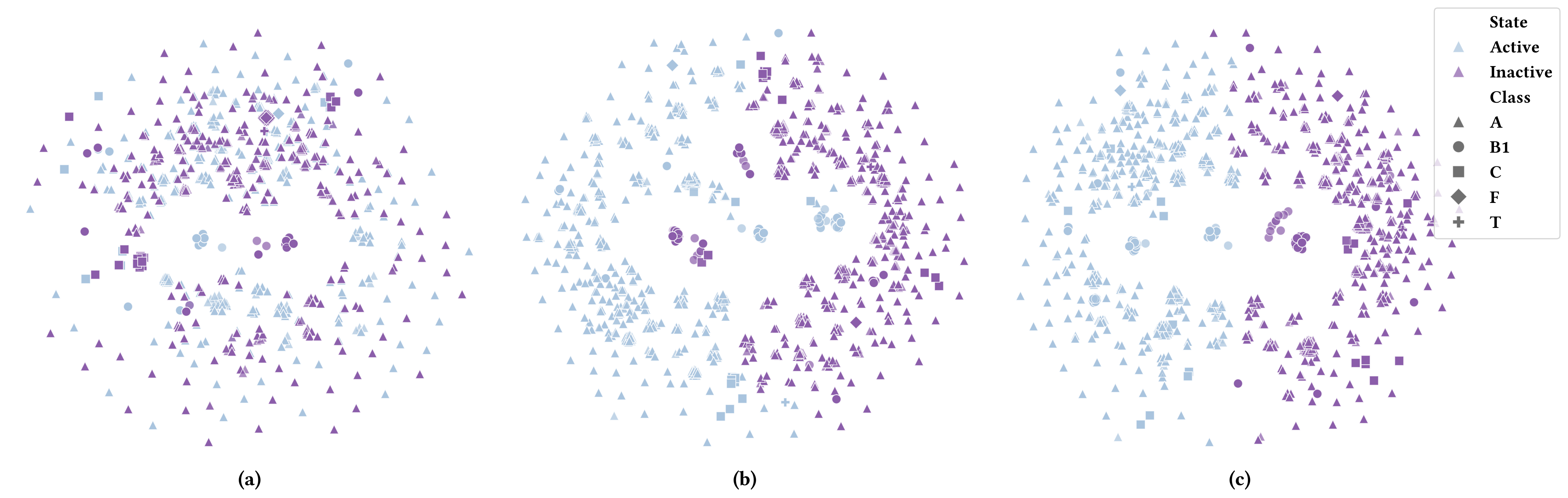}
  \caption{Evolution of latent space disentanglement during training. t-SNE projections of the latent structure views at (a) Epoch 0, (b) Epoch 20, and (c) Epoch 40.  Points are colored by latent state representation type (Blue: Active, Purple: Inactive) and shaped by GPCR family (Triangle: Class A, Circle: Class B1, Square: Class C, Diamond: Class F, Plus: Class T).}
  \label{fig:tsne}
\end{figure*}

To validate that DSQ disentangles receptor conformational states rather than simply memorizing protein identities, we visualized the evolution of latent structural embeddings during training. Figure~\ref{fig:tsne} shows t-SNE projections of the pooled active ($Z_{act}$) and inactive ($Z_{inactive}$) representations at epochs 0, 20, and 40 (convergence). 

\begin{table}[htbp]
\centering
\caption{Quantitative latent-space disentanglement during training. Higher Silhouette Score and lower Davies-Bouldin Index indicate better separation.}
\label{tab:cluster_metrics}
\begin{tabular}{c|cc}
\toprule
Epoch & Silhouette Score$\uparrow$ & Davies-Bouldin Index$\downarrow$ \\
\midrule
0  & 0.0022 & 58.8622 \\
20 & 0.5954 & 0.6585 \\
40 & 0.6086 & 0.6141 \\
\bottomrule
\end{tabular}
\end{table}

\textbf{Emergence of State Disentanglement}. At initialization (Epoch 0, Fig.~\ref{fig:tsne}(a)), the embeddings for active and inactive states overlap greatly. By Epoch 20 (Fig.~\ref{fig:tsne}(b)), separation begins to emerge, and by Epoch 40 (Fig.~\ref{fig:tsne}(c)) the two latent state views form clearer global clusters. This trend is supported quantitatively in Table~\ref{tab:cluster_metrics}: the Silhouette Score increases from 0.0022 to 0.6086, while the Davies-Bouldin Index decreases from 58.8622 to 0.61441.

\textbf{Preservation of Phylogenetic Constraints}. We further analysed the distribution of GPCR families in test set (Class A, B1, C, F, T) within the latent space. As shown in Fig.~\ref{fig:tsne}(a), strong clustering by receptor class is evident even at Epoch 0. This structure is inherited from the pre-trained ESM-2 backbone, which implicitly encodes evolutionary homology. Importantly, this phylogenetic structure is maintained but restructured in Epoch 40. The model preserves protein family features while learning state-specific representations conditional on the protein family.

\section{Conclusion \& Discussions}

In this work, we introduced Dual-State Query (DSQ), a physics-informed deep learning framework for computational GPCR bioactivity profiling. Unlike traditional black-box models, DSQ explicitly simulates the thermodynamic equilibrium between active and inactive receptor conformations by integrating the Monod-Wyman-Changeux (MWC) model with learnable state queries. This architecture allows for simultaneous prediction of binding affinity and functional efficacy while adhering to biophysical constraints. Experiments demonstrate that DSQ improves overall accuracy and ranking performance over the evaluated baselines, with the clearest gains on agonist prediction. Ablation, temperature-sensitivity, perturbation, and clustering analyses support the role of the MWC-inspired inductive bias in shaping the latent activation gate. Homology-stratified evaluation shows that performance remains limited for receptors with low sequence identity to the training set, indicating that protein-language representations alone do not resolve cold-target generalization. Future work may combine the state-aware mechanism with stronger GPCR-specific encoders and structural or conformational information.

Recent deep learning approaches~\cite{ztrk2018deepdta,Nguyen2020graphdta,Shah2025deepdtagen,Velloso2021pdcsmgpcr} use large-scale bioactivity data to predict drug--target interactions. GPCR-specific approaches address additional functional information. AiGPro~\cite{Brahma2025aigpro} performs large-scale GPCR profiling by conditioning bioactivity prediction on an explicit agonist/antagonist class token. Huang et al.~\cite{huang2024decrypting} developed a multitask framework specifically for low-data GPCRs and evaluated it on an independent set of 16 receptors. These approaches address complementary settings. DSQ uses efficacy labels during training but does not require an externally supplied activity-type token at inference; instead, it infers activation probability and bioactivity jointly. Because these studies use different input conditions and evaluation protocols, their published metrics are not directly comparable to the DSQ results. Future work may combine these different settings into a unified framework.

While DSQ offers a physically grounded framework for GPCR profiling, several limitations must be acknowledged. First, the current MWC implementation uses a two-state active/inactive abstraction. This suits the available large-scale labels but cannot explicitly model biased signaling, pathway-specific efficacy, or multiple active conformations. A nature extension is to use $K>2$ state-specific query branches when sufficiently large multi-assay efficacy datasets become available. Second, the present study does not include a receptor-disjoint cold-target test, class-specific predictive evaluation, or an orphan/non-orphan comparison. These evaluations, together with GPCR-specific baselines under a common protocol, are priorities for an extended journal study. Third, DSQ relies on frozen ESM-2 and MolFormer encoders, which may limit adaptation to novel receptor families or chemical scaffolds. Finally, despite improved agonist performance, agonist-antagonist imbalance remains a dataset-level limitation.

\section{GenAI Disclosure}
We used GenAI to polish the writing and fix grammar errors. We carefully reviewed the GenAI outputs to ensure correctness. All scientific methods and experimental results are our own work.

\vspace{1.5em}
{
\small
\setlength{\parskip}{0.8em}

\noindent \textbf{Supplementary Materials:} 

\noindent \textbf{Author Contributions:} Conceptualization, Shuo Zhang; methodology, Shuo Zhang; software, Shuo Zhang; validation, Shuo Zhang; formal analysis, Shuo Zhang; investigation, Shuo Zhang; resources, Shuo Zhang; visualization, Shuo Zhang; data curation, Shuo Zhang, Huifeng Zhang, Rongqi Hong;  writing---original draft preparation, Shuo Zhang, Huifeng Zhang, Rongqi Hong, Jian K. Liu; writing---review and editing, Shuo Zhang, Huifeng Zhang, Rongqi Hong, Jian K. Liu; supervision, Jian K. Liu.; project administration, Jian K. Liu.

\noindent \textbf{Funding:} This work was partially supported by BlueBEAR and Baskerville, funded by the EPSRC and UKRI through the World Class Labs scheme (EP/T022221/1) and the Digital Research Infrastructure programme (EP/W032244/1) and operated by Advanced Research Computing at the University of Birmingham.

\noindent \textbf{Data Availability Statement:} The dataset, trained model weights, and code are available at \url{https://github.com/jiankliu/DSQ}.

\noindent \textbf{Conflicts of Interest:} The authors declare no conflict of interest. The funders had no role in the design of the study; in the collection, analyses, or interpretation of data; in the writing of the manuscript; or in the decision to publish the results.
\par
}

\bibliographystyle{unsrt}
\bibliography{reference}

\newpage
\setcounter{figure}{0}
\renewcommand{\thefigure}{S\arabic{figure}}
\begin{figure*}[h]
    \centering
    \includegraphics[width=0.8\textwidth]{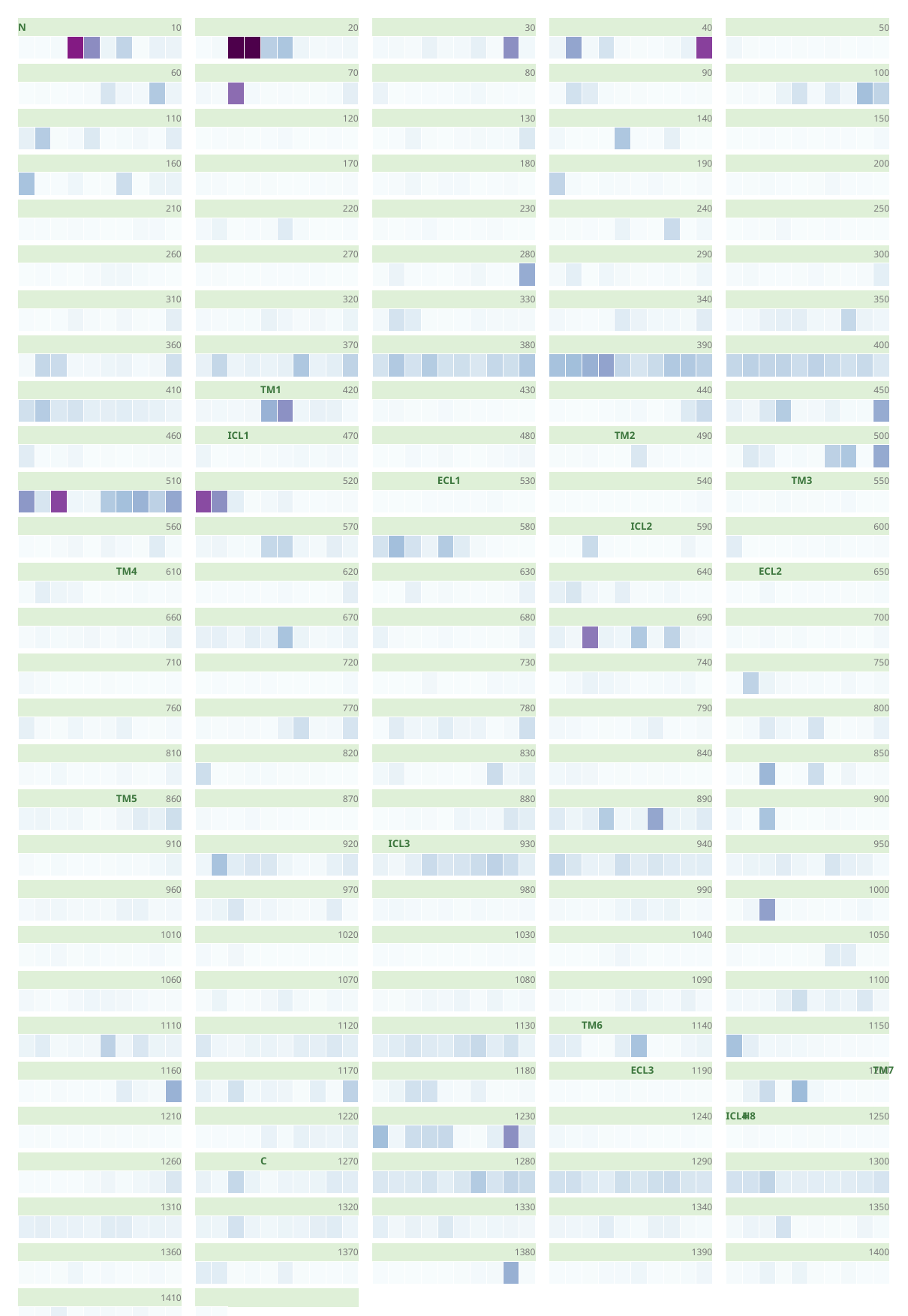}
    \caption{Differential attention weights of class A receptors in test set}
    \label{fig:attn_weight_classA_heatmap}
\end{figure*}
\begin{figure*}[h]
    \centering
    \includegraphics[width=0.8\textwidth]{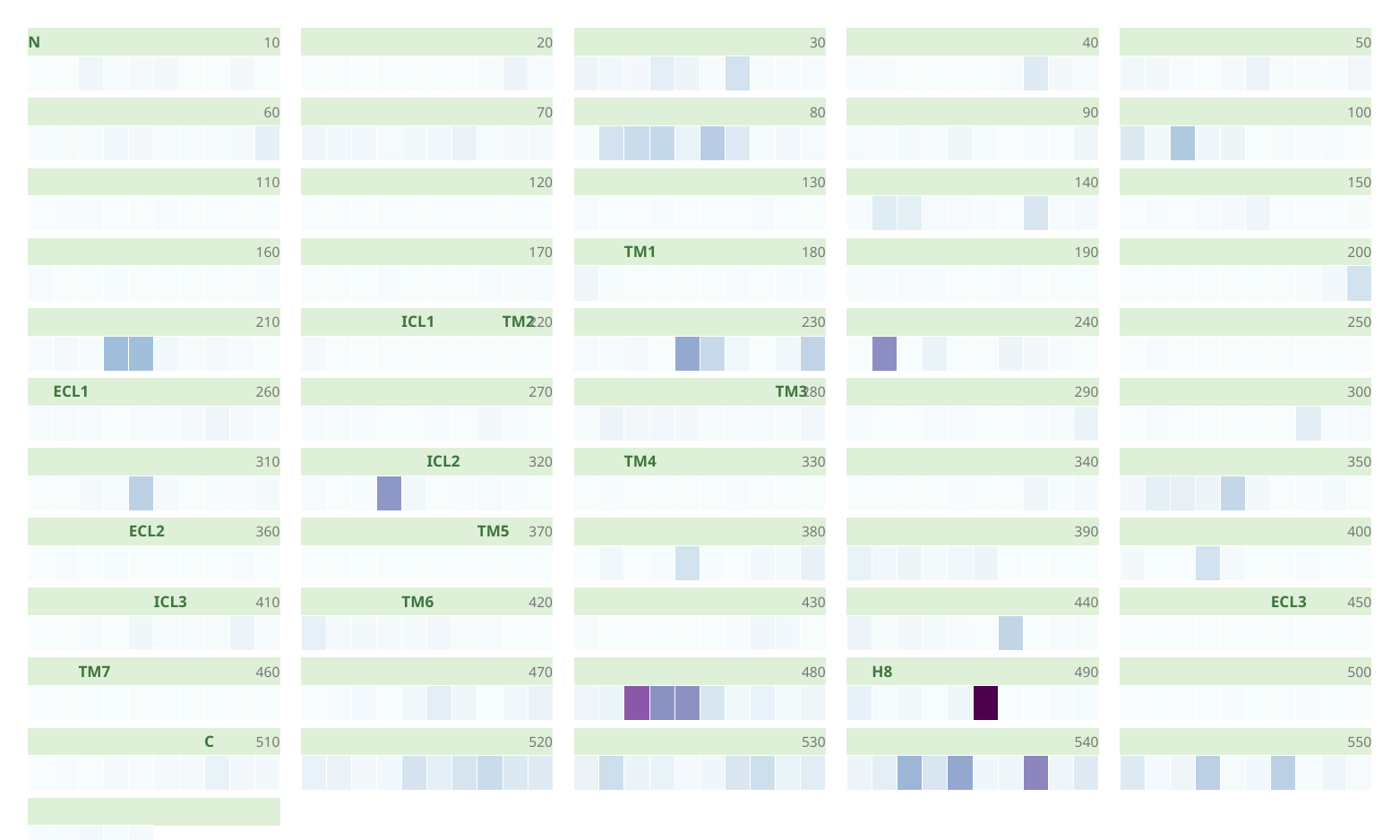}
    \caption{Differential attention weights of class B1 receptors in test set}
    \label{fig:attn_weight_classB1_heatmap}
\end{figure*}

\end{document}